\documentstyle[12pt]{article}
\def\beq{\begin{equation}}
\def\eeq{\end{equation}}
\def\beqa{\begin{eqnarray}}
\def\eeqa{\end{eqnarray}}

\begin{document}
\begin{flushright}
VAND--TH-96--6 \\
UH--511--873--97 \\
January 1998
\end{flushright}
\begin{center}
\Large
{\bf The Low Energy Behavior of some Models with
Dynamical Supersymmetry Breaking}
\end{center}
\normalsize
\bigskip
\begin{center}
{\large  Tonnis A. ter Veldhuis
}
\\
{\sl Department of Physics \& Astronomy\\
University of Hawaii\\
Honolulu, HI 96822, USA}\\
\vspace{2cm}
\end{center}
\centerline{ABSTRACT}
\vspace{1cm}
\noindent
We study supersymmetric $SU(5)$ chiral gauge theories 
with 2 fields in the $10$ representation,  
$2+N_F$ fields in the $\bar{5}$ representation 
and $N_F$ fields in the $5$ representation, 
for $N_F=0,1,2$. 
With a suitable superpotential,
supersymmetry is shown to be broken dynamically for each 
of these values of $N_F$.
We analyze the calculable limit for the model with $N_F=0$ 
in detail, and determine the low energy effective sigma model in this case.
For $N_F=1$ we find the quantum moduli space, and for $N_F=2$ we construct 
the s--confining potential.
\\
\vfill\eject

\section*{Introduction and Results}

In this Letter we study the low energy dynamics of
supersymmetric $SU(5)$ chiral gauge theories with two matter fields
$T_{a}^{ij}$ in the $10$ representation,  $2+N_F$ fields $\bar{F}_i^{\alpha}$
in the $\bar{5}$ representation  and $N_F$ fields $F_{\sigma}^i$ 
in the $5$ representation, for $N_F =0,1$ and $2$. 
Here $i,j=1..5$ are $SU(5)$ indices 
and $\alpha$, $a$ and $\sigma$ are flavor indices.
With this matter content, the theories are anomaly free. 
They are also asymptotically free, as the one loop
coefficient of the beta function is given by $b_0 = 11 - N_F$.  
For each value of $N_F$, we describe the classical moduli
space  in terms
of holomorphic gauge invariant polynomials \cite{ADS1,LT,PR}.
The manner in which the quantum moduli space differs from the classical
one
depends on the value of $N_F$; 
for $N_F=0$ the classical moduli space is completely lifted by
a dynamically generated superpotential, for $N_F=1$ one
of the constraints among the gauge invariants is modified, 
and for $N_F=2$ the quantum moduli
space is identical to the classical one.
The low energy behavior for $N_F=0$, $1$ and $2$ is therefore similar to
supersymmetric QCD with $N_C$ colors and $N_C-1$, $N_C$ and $N_C+1$ flavors,
respectively \cite{IS,Pes,Shi}. However, in contrast to supersymmetric
QCD, in the models we consider supersymmetry is broken 
dynamically when suitable tree level superpotentials are added.
For each value of $N_F$, the physics below the dynamical scale 
of the gauge interactions can be described by an
effective theory. The light degrees of freedom in the effective
theory are adequately represented by the same gauge invariants
that characterize the moduli space. 
We construct these effective theories with the objective to show
that supersymmetry is broken.  
This method of analysis is
consistent as long as the supersymmetry breaking scale is below 
the dynamical scale. 
In general, the K\"{a}hler potential is under much less
control than the superpotential. (The calculable limit of
the model for $N_F=0$ is an exception.)
However, when the low energy degrees of freedom have been identified
correctly, the K\"{a}hler potential is supposedly well--behaved.
Under this assumption, it is possible  to draw qualitative conclusions
from an analysis of the superpotential alone. 
Finally, we show how the models for different values
of $N_F$ are related by holomorphic decoupling.

For $N_F = 0$ the well-known $SU(5)$ model \cite{ADS2,TtV}
with calculable dynamical supersymmetry breaking is obtained.
In this model the classical moduli space is completely lifted
by a dynamically generated superpotential. 
Supersymmetry is broken when a tree level potential 
is introduced which prevents the vacuum from moving to infinity.
When the coupling constant of the tree level superpotential is 
sufficiently small,
the vacuum occurs at weak coupling, with expectation values of the 
fields which lie near the
classical moduli space and which are
much larger than the dynamical scale of the theory. 
The model is thus seen to be very similar to the
$SU(3) \otimes SU(2)$ \cite{ADS1,BPR,IT} model, the paradigm of models
with calculable dynamical supersymmetry breaking. 

We discuss the calculable limit of the $N_F=0$ model from three
different angles. In Section 1 we review the model and we 
extend our previous numerical calculation \cite{TtV} of the mass spectrum
to next to leading order in the ratio of the superpotential
coupling and the gauge coupling.
In Section 1.1 we construct the low energy effective sigma model
in terms of the moduli fields, including both the K\"{a}hler potential
and the superpotential.
With this sigma model in hand, we recalculate the vacuum energy,
vacuum expectation values and the mass spectrum.
The global symmetries of the sigma model elucidate the origin of
degeneracies in the mass spectrum which appeared accidental
in the full theory \cite{TtV}.
In Section 1.2 we give an explicit parametrization in terms
of vacuum expectation values of the fundamental fields of a
particular direction in the classical moduli space which includes the 
vacuum. This result provides yet another way to calculate
quantities such as the mass spectrum, vacuum energy and expectation
values. 

In Section 2 we add a flavor to the model. For $N_F = 1$ the classical
moduli space is modified by nonperturbative effects, and
the classical singularities are removed. We add a suitable
superpotential with a mass term for the additional flavor and a
Yukawa term. With the additional assumption
that the K\"{a}hler potential is non--singular, we then
show that supersymmetry is broken. We also check
the consistency of our results; when 
the additional flavor is integrated out by 
taking its mass much larger than the dynamical
scale, 
the theory for $N_F=0$ with the correct superpotential ensues.

In Section 3 we add yet another flavor to the theory.
In the case $N_F=2$ there are no modifications to the 
classical moduli space. 
However, the singularities of the moduli space
are interpreted differently. At the classical level singularities
in the moduli space indicate that the gauge
symmetry is only partially broken, and as a consequence the 
spectrum contains additional massless
gauge multiplets at these points. At the quantum level, the theory confines, 
and the singularities are associated with additional massless 
composite degrees of freedom.
We determine the confining superpotential.  When 
a tree level superpotential with mass terms
for the two flavors and a Yukawa term is added to the model, supersymmetry
is shown to be  broken. We also show that when one of the flavors
is integrated out,
the theory with $N_F = 1$ with the correct quantum
modified moduli space results.

\section{$N_F = 0$}
 
% - the model; overview; what has been done.

In the absence of a superpotential 
the model with $N_F=0$ has an 
$SU(2)_T \otimes SU(2)_{\bar{F}} \otimes U(1)_{A'} \otimes U(1)_{R'}$ 
global symmetry, under which the fundamental fields transform as
\begin{center}
\begin{tabular}{|l|c|c|c|c|}
\hline
                    & $SU(2)_{\bar{F}}$ & $SU(2)_T$ & $U(1)_{A'}$ & $U(1)_{R'}$  \\ 
\hline
$\bar{F}^{\alpha}$  &   $2$             &   $1$     &   $ 3/5$      &   $-4$       \\
$T_a$               &   $1$             &   $2$     &   $-1/5$      &   $1$        \\
\hline 
\end{tabular}\ \ .
\end{center}
The classical
moduli space is parametrized by the expectation values of
the six holomorphic gauge invariant
polynomials\footnote{These invariants are slightly different from the ones
defined in Ref.\cite{TtV} in order to make  their transformation properties explicit.}
\beqa
X_a & = & \epsilon_{\alpha \beta} \bar{F}_i^{\alpha} 
\bar{F}_j^{\beta} T_a^{ij},
\nonumber \\
J_a^{\alpha} & = & \epsilon_{ijklm} \bar{F}_n^{\alpha}
 T_a^{ij}  T_b^{kl} T_c^{mn} \epsilon^{bc}. \label{mf}
\eeqa
Under the
global symmetry transformations these gauge invariants transform as
\begin{center}
\begin{tabular}{|l|c|c|c|c|}
\hline
                    & $SU(2)_{\bar{F}}$ & $SU(2)_T$ & $U(1)_{A'}$ & $U(1)_{R'}$  \\ 
\hline
$X_a $              &   $1$             &   $2$     &   $ 1$      &   $-7$       \\
$J_a^{\alpha} $     &   $2$             &   $2$     &   $ 0$      &   $-1$       \\
\hline 
\end{tabular}\ \ .
\end{center}
Holomorphy and symmetries uniquely determine the form of
a possible dynamically generated
superpotential\footnote{The sole purpose of the factor $9/2$ here 
is to maintain 
consistency with Ref.\cite{TtV}.}
\beq
W_{np} = \frac{9}{2} \frac{\Lambda_0^{11}}{
J_a^{\alpha} J_b^{\beta}  \epsilon_{\alpha \beta} \epsilon^{ab}}. 
\label{Wnp}
\eeq
The scale $\Lambda_0$ can be defined by relating it
to $\overline{\Lambda_0}$, the scale at which the one--loop renormalization
group equation
for the gauge coupling diverges. In principle, this
requires an instanton calculation,
but we can make use of the fact that in some D--flat
directions the model reduces to a simpler model for which such
an instanton calculation has been performed already. 
In the D-flat direction with $T_1^{45} = -T_1^{54}= \bar{F}_4^1 = \bar{F}_5^2=a$, 
and all other components vanishing,
the $SU(5)$ gauge symmetry is broken to $SU(3)$.
Apart from two singlets, the theory below the
scale $a$
is  supersymmetric QCD with three colors and two flavors. 
Expanding the superpotential (\ref{Wnp}) around this flat
direction yields
\beq
W_{np} = \frac{1}{64} \frac{\Lambda_0^{11}}{a^4 Det(\bar{Q}^p Q^q)}. 
\label{sp1}
\eeq
Here the indices $p$ and $q$ label the flavor of the quarks $Q$.
The dynamically generated superpotential 
for the $SU(3)$  theory has been calculated \cite{FP} to be
\beq
W_{(3)} = \frac{\overline{\Lambda}_{(3)}^7}{Det(\bar{Q}^p Q^q)},
\label{sp2}
\eeq
where $\overline{\Lambda}_{(3)}$ is the scale at which the
$SU(3)$ gauge coupling diverges.
A comparison of the superpotentials (\ref{sp1}) and (\ref{sp2}) yields 
$\Lambda_{0}^{11} =  64 a^4 \overline{\Lambda}_{(3)}^7$.
Matching the $SU(3)$ and $SU(5)$ gauge coupling at the scale
$a$ results in 
$\overline{\Lambda}_0^{11} = a^4 \overline{\Lambda}_{(3)}^7$.
As a consequence  $\Lambda_0 = 2^{6/11} \overline{\Lambda}_0$.

At tree level a renormalizable superpotential
can be added to the model, so that the full superpotential is
\beq
W = W_{np} + \lambda X_1 . \label{superpotential}
\eeq
The renormalizable part of the superpotential explicitly breaks the $SU(2)_T$ 
flavor symmetry, but 
a non-anomalous $SU(2)_{\bar{F}} \otimes U(1)_A \otimes U(1)_R$ symmetry
remains. Under this
reduced symmetry group the fundamental and composite fields transform as
\begin{center}
\begin{tabular}{|l|c|c|c|c|}
\hline
                    & $SU(2)_{\bar{F}}$ & $U(1)_{A}$   & $U(1)_{R}$  \\ 
\hline
$\bar{F}^{\alpha}$  &   $2$             &   $ 3/5$      &   $-4$       \\
$T_1$               &   $1$             &   $-6/5$      &   $10$       \\
$T_2$               &   $1$             &   $ 4/5$      &   $-8$       \\
\hline \hline 
$X_1$               &   $1$             &   $ 0  $      &   $2$        \\
$X_2$               &   $1$             &   $ 2  $      &   $-16$      \\
$J_1^{\alpha}$      &   $2$             &   $-1  $      &   $8$        \\
$J_2^{\alpha}$      &   $2$             &   $ 1  $      &   $-10$      \\
\hline 
\end{tabular}\ \ .
\end{center}
For small values of $\lambda$
the $SU(5)$ gauge symmetry is completely broken by vacuum expectation
values which are much larger than the dynamical scale of
the gauge interactions. In earlier work\cite{TtV} the
particle spectrum was calculated in this case
by numerically minimizing the scalar potential of the model.
The vacuum energy was also calculated and found to be non-vanishing.
Hence it was verified that supersymmetry is broken in this
model.
In addition, it was determined that the global internal 
symmetries are broken to a single remaining
$U(1)_Q$ with $Q=I_{(\bar{F})}^3 + A/2$. 
When the value of $\lambda$ is increased, the vacuum expectation values
of the fields approach the dynamical scale. 
The vacuum becomes strongly interacting and the K\"{a}hler
potential is no longer under control. 
However, there appears to be  no invariant distinction
between a Higgs description and a confined description. Supersymmetry
is therefore likely also broken for large values of $\lambda$,
although the model is then not \mbox{\lq\lq calculable\rq\rq}.
In addition, the moduli fields are the appropriate degrees
of freedom for the low energy effective
theory for both small and large values of $\lambda$.

%Supersymmetry is seen to be broken because $(X^1)_F^{\dagger}=\lambda \neq 0$.

The model has only two parameters, which, for example, can 
be chosen as the dynamical
scale $\Lambda_0$ and the scale $v$ of the vacuum expectation values
of the scalar fields. In order to quantify
for what range of the parameters the model is perturbative, note
that the coupling constant $\lambda$ at the scale $v$ varies as
\beq
\lambda(v) \sim \left( \frac{\Lambda_0}{v} \right)^{11}, \label{sclambda}
\eeq
whereas the gauge coupling $g$ scales as
\beq
g^2(v) \sim - \frac{8 \pi^2}{11 \ln \frac{\Lambda_0}{v}} \label{scg}.
\eeq
The model is calculable if the gauge coupling is in the perturbative
range at the scale $v$. Taking this to mean  that $g^2(v) < 8 \pi^2$, 
it follows from Eq.(\ref{scg}) that $v >  e^{\frac{1}{11}}\Lambda_0$.
Eq.(\ref{sclambda}) then implies that $\lambda< e^{-1}$.

In the calculable limit $g(v)$ is much larger than $\lambda(v)$. 
As a consequence, a sequence of scales arises. In decreasing order
there is a large mass scale $\sim g v$, the scale of the vacuum
expectation values of the scalar components $v$, the
dynamical scale $\Lambda_0$, the supersymmetry breaking scale
$\sim \lambda^{1/2} v $, and a light mass scale $\lambda v$. 
The heavy sector of the spectrum contains the components of
twenty four massive vector multiplets. The light spectrum contains
the components of the six chiral multiplets which remain
after the other chiral multiplets are absorbed by the vector multiplets
through the Higgs mechanism.
The light spectrum includes four massless scalars and two 
massless fermions. The massless scalars are the Goldstone bosons associated
with the spontaneously broken global internal symmetries. One of
the massless fermions is a Goldstino associated with
the spontaneous breaking of global supersymmetry. The second massless
fermion is charged and saturates the anomaly matching condition for the 
unbroken $U(1)_Q$ symmetry.
The masses of the light particles can be
expanded in powers of $\lambda/g$ as follows:
\beq
m =  \lambda^{10/11}  \Lambda_0 
\left( \mu_0 + \frac{1}{2} r \left( \frac{\lambda}{g} \right)^2 + 
{\cal O} \left(\frac{\lambda}{g}\right)^4  \right). \label{meq}
\eeq
Previously\cite{TtV}, the masses of the
light scalars and fermions were calculated
in leading order in $\lambda/g$.
In that calculation the full scalar potential of the model, including
D-- and F--terms, was minimized numerically, and the masses of both
the light and heavy particles 
were obtained by expanding the theory
around the minimum. We have now extended this calculation to 
next to leading order in $\lambda/g$.
The resulting values for the parameters $\mu_0$ and $r$ 
are listed in Table \ref{smass}
for scalars and in Table \ref{fmass} for fermions. 
%Note
%that $g(v)$ is not an independent parameter, but a function of
%$\lambda(v)$ and $\Lambda_0$.
As was already observed in Ref.\cite{TtV}, degeneracies in the values of 
$\mu_0$  indicate that
to lowest order in  $\lambda/g$ 
the spectrum displays
an $SU(2)_D \otimes U(1)_{A_e}$ global symmetry. 
This group is not a subgroup of the global symmetry group
of the model.
Moreover, its transformations do not consist of combinations
of transformations from the gauge group (with constant parameters)
and the global symmetry group either. We will
show how the moduli fields transform when we discuss the 
low energy effective sigma model in the next section.
$SU(2)_D \otimes U(1)_{A_e}$ is broken to $U(1)_Q$ with $Q = I_D^3+A_e$
in next to leading order in  $\lambda/g$,
as follows from the values of $r$.

\begin{table}
\begin{center}
\begin{tabular}{|c|c||c||l|l|}
\hline
  $SU(2)_D$    &   $A_e$     &  $Q$       & $\mu_0$   & $r$     \\ 
\hline
   $1$         &   $0$       &  $0$       &  $0$      & $0$     \\
\hline
               &             &  $-1$      &           & $0$     \\
   $3$         &   $0$       &  $+1$      &  $0$      &         \\
\cline{3-3} \cline{5-5}
               &             &  $0$       &           & $0$     \\
\hline
   $1$         &   $-1$      &  $-1$      &  $2.550$  & $-44.9$ \\
               &   $+1$      &  $+1$      &           &         \\
\hline
               &             &  $-1$      &           & $-12.3$  \\
   $3$         &   $0$       &  $+1$      &  $2.744$  &         \\
\cline{3-3} \cline{5-5}
               &             &  $0$       &           & $-20.6$  \\
\hline
   $1$         &   $0$       &  $0$       &  $3.904$  & $-66.4$ \\
\hline
   $1$         &   $0$       &  $0$       &  $5.946$  & $-2.3$  \\
\hline
   $1$         &   $0$       &  $0$       &  $9.320$  & $-45.7$  \\
\hline 
\end{tabular}
\end{center}
\caption{The parameters $\mu_0$ and $r$ which determine
the twelve scalar masses according to Eq.\protect{(\ref{meq})}. Each scalar is
classified according to its $SU(2)_D \otimes U(1)_{A_e}$ representation 
and $U(1)_Q$ charge.} 
\label{smass}
\end{table}

\begin{table}
\begin{center}
\begin{tabular}{|c|c||c||l|l|}
\hline
  $SU(2)_D$    &   $A_e$     &  $Q$       & $\mu_0$   & $r$    \\ 
\hline
   $1$         &   $0$       &  $0$       &  $0$      & $0$    \\
\hline
   $1$         &   $+1$      &  $+1$      &  $0$      & $0$    \\
\hline
               &             &  $-1$      &           & $-3.7$ \\
   $3$         &   $0$       &  $+1$      &  $0.716$  &        \\
\cline{3-3} \cline{5-5}
               &             &  $ 0$      &           & $+0.1$ \\
\hline
   $1$         &   $0$       &  $ 0$      &  $7.486$  & $-18.7$ \\
\hline 
\end{tabular}
\end{center}
\caption{The parameters $\mu_0$ and $r$ which determine
the six fermion masses according to Eq.\protect{(\ref{meq})}. Each fermion is
classified according to its $SU(2)_D \otimes U(1)_{A_e}$ representation 
and $U(1)_Q$ charge.} 
\label{fmass}
\end{table}

%- outline of the sections with conclusions
  
\subsection{The low energy effective sigma model}

In the calculable limit the vacuum expectation values are
large and in the vicinity of the D--flat directions. 
Below the dynamical scale $\Lambda_0$ the heavy vector
multiplets can be integrated out. 
The interactions of
the remaining light degrees of freedom are adequately described
by a sigma model with the
moduli fields $X_{a}$ and $J_{a}^{\alpha}$ as coordinates.
The K\"{a}hler potential of this sigma model is the classical
K\"{a}hler potential on the moduli space.
This K\"{a}hler potential can be determined by
the method of Affleck, Dine and Seiberg\cite{ADS1}. According
to this method, 
the classical K\"{a}hler potential of the fundamental theory 
with the gauge interactions switched off is projected onto the moduli fields.
The method is powerful because it is non--local. This means that it
is not necessary to specify a specific point in the moduli
space around which the theory is expanded. As a consequence it
is not necessary to minimize the scalar potential of the fundamental
theory to find the vacuum expectation values of the moduli fields.
The construction of the sigma model is completed when the
non--perturbative and tree level superpotentials in terms
of the moduli fields are added.

The effective sigma model is supersymmetric and invariant under the
full global symmetry group of the fundamental theory. However, when
its scalar potential is minimized some of the moduli fields get
expectation values. Supersymmetry and some of the global symmetries
are then spontaneously broken. 
The sigma model is thus equivalent to the fundamental theory
after the heavy degrees of freedom are integrated out.

Applying this method, the effective  K\"{a}hler potential of the sigma model
is invariant under all internal global symmetries and
supersymmetry. Hence it takes the form
\beq
K_{eff}=K_{eff}(I_1,I_2,I_3,I_4),
\eeq
with
\beqa
I_1 & = & {{X}^a}^{\dagger} X_a  \nonumber \\
I_2 & = & {J_{\alpha}^a}^{\dagger} J_a^{\alpha}  \nonumber\\
I_3 & = & {X^a}^{\dagger} {J_{\beta}^b}^{\dagger} X_b J_a^{\beta} \nonumber \\
I_4 & = & {J_{\alpha}^a}^{\dagger} {J_{\beta}^b}^{\dagger} J_a^{\beta} J_b^{\alpha}.
\eeqa
Following the Affleck, Dine and Seiberg procedure, the
objects ${T^a}^{\dagger} T_a$ and $\bar{F}_{\alpha}^{\dagger}
\bar{F}^{\alpha}$ have to be related to
$I_1$, $I_2$, $I_3$ and $I_4$ in the moduli space.
Defining
\beqa
A & = & 125 I_1 \nonumber \\
B & = & \frac{25}{9} 
\left( \sqrt{\frac{1}{2} I_2 + \frac{1}{2} \sqrt{2 I_4 - I_2^2}} +
       \sqrt{\frac{1}{2} I_2 - \frac{1}{2} \sqrt{2 I_4 - I_2^2}} \right),
\eeqa
we proved that the relations
\beqa
A & = & ({T^a}^{\dagger} T_a + 4 \bar{F}_{\alpha}^{\dagger} \bar{F}^{\alpha})^2
(-\frac{1}{2} {T^a}^{\dagger} T_a + 3 \bar{F}_{\alpha}^\dagger \bar{F}^{\alpha}) \nonumber \\
B & = & \frac{2}{3} ({T^a}^{\dagger} T_a + 4 \bar{F}_{\alpha}^{\dagger} \bar{F}^{\alpha}) 
({T^a}^{\dagger} T_a - \bar{F}_{\alpha}^{\dagger} \bar{F}^{\alpha}) \label{syseq}
\eeqa
are identically valid in the D--flat directions.
The K\"{a}hler potential of the sigma model 
is  determined by inverting these identities in order to
write the K\"{a}hler potential of the fundamental
theory
in terms of  the moduli fields. Defining
$p={T^a}^{\dagger} T_a + 4 \bar{F}_{\alpha}^{\dagger} \bar{F}^{\alpha}$
and $q={T^a}^{\dagger} T_a -  \bar{F}_{\alpha}^{\dagger} \bar{F}^{\alpha}$,
the system of equations (\ref{syseq})
is equivalent to 
\beqa
p^3 - 3 B p - 2 A & = & 0 \nonumber \\
q & = & \frac{3}{2} \frac{B}{p}.
\eeqa
Solving the cubic equation for $p$ and 
choosing the appropriate root gives
\beq
p = 2 \sqrt{B} 
\cos ( \frac{1}{3} \arccos \frac{A}{B^{\frac{3}{2}}} ).
\eeq
The resulting effective K\"{a}hler potential in 
terms of $I_1$, $I_2$, $I_3$ and $I_4$ is
\beq
K_{eff}  = 
(\frac{1}{2}{T^a}^{\dagger} T_a + \bar{F}_{\alpha}^{\dagger} \bar{F}^{\alpha} ) |_{flat}
= \frac{1}{10} \left( 3p + 2 q \right)
= \frac{3}{10} \left( p + \frac{B}{p} \right). \label{kp}
\eeq
Surprisingly, $K_{eff}$ does not depend on $I_3$. As a consequence,
the global symmetries of the effective K\"{a}hler potential extend
those of the underlying fundamental theory. 
The complete global symmetry group of the effective K\"{a}hler 
potential is 
$SU(2)_{\bar{F}} \otimes SU(2)_{1} \otimes SU(2)_{2} \otimes U(1)_{R'}$,
under which the moduli fields transform as 
\begin{center}
\begin{tabular}{|l|c|c|c|c|}
\hline
                & $SU(2)_{\bar{F}}$ & $SU(2)_{1}$ &   $SU(2)_{2}$   & $U(1)_{R'}$  \\ 
\hline
$X_b$           &   $1$             &   $1$        &   $ 2$           &   $-4$       \\
$J_a^{\alpha}$  &   $2$             &   $2$        &   $ 1$           &   $1$        \\
\hline      
\end{tabular}\ \ .
\end{center}
To keep track of the action of the various $SU(2)$ symmetry transformations,
note that $a$ is an $SU(2)_1$ index, $b$ is an $SU(2)_2$ index, and
$\alpha$ is an $SU(2)_{\bar{F}}$ index as before.
The original $SU(2)_T$
group is the vectorial subgroup of $ SU(2)_{1} \otimes SU(2)_{2} $.
In addition, observe that because the moduli fields are invariant under gauge
transformations, $SU(2)_1$ and $SU(2)_2$ are not just combinations
of gauge transformations with a constant parameter and $SU(2)_T$ 
transformations.
All information pertaining to the low energy limit of the $N_F=0$
model is contained in the sigma model defined by the K\"{a}hler
potential (\ref{kp}) and the superpotential (\ref{superpotential}).
The tree level superpotential  breaks some of the global symmetries
explicitly. The remaining symmetry group is
$SU(2)_F \otimes SU(2)_{1} \otimes U(1)_{Ae} \otimes U(1)_{Re}$. Under
this group the moduli fields transform as
\begin{center}
\begin{tabular}{|l|c|c|c|c|}
\hline
                & $SU(2)_{\bar{F}}$ & $SU(2)_{1}$  &  $U(1)_{Ae}$  & $U(1)_{Re}$ \\ 
\hline
$X_1$           &   $1$     &  $ 1$         &   $0$         &   $2$      \\
$X_2$           &   $1$     &  $ 1$         &   $1$         &   $0$      \\
$J_a^{\alpha}$  &   $2$     &  $ 2$         &   $0$         &   $-1$     \\
\hline   
\end{tabular}\ \ .
\end{center}
Taking into account the non-canonical form of the K\"{a}hler
potential, the scalar potential of the sigma model was minimized. The
resulting vacuum energy was found to be
\beq
V= 2.807 \lambda^{18/11} \Lambda_0^4.
\eeq
The vacuum expectation values of the moduli fields in the
minimum were found to be
\beqa
X_1 & = & 0.247 \lambda^{-3/11} \Lambda_0^3 \nonumber \\
X_2 & = & 0 \nonumber \\
J_1^1 & = & - 3.452 \lambda^{-4/11} \Lambda_0^4 \nonumber \\
J_2^1 & = & 0 \nonumber \\
J_1^2 & = & 0 \nonumber \\
J_2^2 & = & - 3.452 \lambda^{-4/11} \Lambda_0^4, \label{vevs}
\eeqa
in accordance with the calculation in the full theory \cite{TtV}.
The form of the vacuum expectation values is such that both
the order parameter for  $U(1)_Q$ symmetry breaking,
$I_3^2-I_1 I_2 I_3 + 1/2 I_1^2 I_2^2 - 1/2 I_1^2 I_4$,
and the order parameter for $SU(2)_D$ breaking,
$2 I_4 - I_2^2 $,
vanish. This is not surprising, since points with extended symmetry
are stationary points of the potential.
The symmetry breaking pattern is thus
$SU(2)_{\bar{F}} \otimes SU(2)_{1} \otimes  U(1)_{Ae}  \otimes U(1)_{Re}
\rightarrow SU(2)_D \otimes U(1)_{Ae}$, where $SU(2)_D$ is
a diagonal subgroup of $SU(2)_{\bar{F}} \otimes SU(2)_1$.
In order to discuss the mass spectrum, it is useful to
study fields that transform under irreducible representations
of the unbroken subgroup.
The six fields $X_1$, $X_2$, $N=J_{a}^{\alpha} \delta_{\alpha}^{a}$
and $S^{i} = J_{a}^{\alpha} (\sigma^i)_{\alpha}^a$ with transformation
properties
\begin{center}
\begin{tabular}{|l|c|c|c|c|}
\hline
                & $SU(2)_D$ & $U(1)_{A}$  \\ \hline
$X_1$            &   $1$     &   $0$       \\
$X_2$            &   $1$     &   $1$       \\
$N$               &   $1$     &   $0$       \\
$S^i$             &   $3$     &   $0$       \\
\hline   
\end{tabular}\ \ ,
\end{center}
form an equivalent coordinate system for the sigma model.
All masses and interactions of the theory can be determined
by expanding the K\"{a}hler potential and the superpotential
around the vacuum expectation values (\ref{vevs}).
We calculated the scalar and fermion masses in the sigma model, 
and the results were in complete agreement with the light masses 
determined in the full theory to lowest order of 
$\lambda/g$, as they can be gleaned from
Eq.(\ref{meq}) and
Tables \ref{smass} and \ref{fmass}.  
The  Goldstino is a linear
combination of the fermionic components of $X_1$ and $N$. The
charged massless fermion forms the fermionic component of $X_2$,
while the scalar components of this field correspond to the massive
complex charged scalar in the spectrum.
The scalar components of $S^i$ form a massless and a massive
triplet of real scalars, and one linear combination of the scalar components
of $X_1$ and $N$ corresponds to the neutral massless scalar.

\subsection{Parametrization of the moduli space}

The classical moduli space is described by twelve parameters, 
which  can be conveniently
chosen as the six complex vacuum expectation values of the gauge invariants
(\ref{mf}). Alternatively, the moduli space can be described
by the vacuum expectation values of the fundamental fields.
The D--flat directions are given by the solutions
to the equation \cite{ADS1}
\beq
T_{ij}^{a\dagger} T_{a}^{ik} - \bar{F}_{\alpha}^{k\dagger} \bar{F}_j^{\alpha} \sim
{\delta}_j^k. \label{flateq}
\eeq 
modulo a gauge transformation.
The moduli space is  determined by a four parameter solution
to Eq.(\ref{flateq}) which breaks all global symmetries, and the
eight parameters of global 
$SU(2)_{\bar{F}} \otimes SU(2)_T \otimes U(1)_{A'} \otimes U(1)_{R'}$ 
transformations. We did  construct a generic four parameter solution,
and we used it to check the identities (\ref{syseq}) in the previous
section.
However, as the minimum of the scalar
potential occurs in a special direction of the moduli space in which the
$SU(2)_D$ symmetry is not broken, we just provide a parametrization
of  this special direction here: 
\begin{eqnarray}
T_{1}=\left(
\begin{array}{ccccc}
0 & a & 0 & e & 0 \\
-a & 0 & p & 0 & 0 \\
0 & -p & 0 & q & 0 \\
-e & 0 & -q & 0 & 0 \\
0 & 0 & 0 & 0 & 0 
\end{array}
\right)
T_{2}=\left(
\begin{array}{ccccc}
0 & 0 & 0 & 0 & r \\
0 & 0 & 0 & b & 0 \\
0 & 0 & 0 & 0 & s \\
0 & -b & 0 & 0 & 0 \\
-r & 0 & -s & 0 & 0 
\end{array}
\right)
\bar{F}^{1}=
\left(
\begin{array}{c}
m \\
0 \\
0 \\
0 \\
0
\end{array}
\right)
\bar{F}^{2}=
\left(
\begin{array}{c}
0 \\
0 \\
0 \\
n \\
0
\end{array}
\right).
\eeqa
The independent parameters of this solution are $a$ and $b$. The remaining
parameters in terms of $a$ and $b$ are
\beqa
q & = & \sqrt{a^2 + a b} \nonumber \\
e & = & (b+a) \sqrt{\frac{b-a}{a}}  \nonumber \\
r & = & \sqrt{b^2 + a b} \nonumber \\
m & = & b\sqrt{\frac{b}{a}}  \nonumber \\
n & = & b\sqrt{\frac{b}{a}}\nonumber \\
s & = & -\sqrt{b^2 - a b} \nonumber \\
p & = & \sqrt{b^2 - a^2}.
\eeqa
The expectation values of the gauge invariant polynomials in 
this flat direction are
\beqa
X_1 & = & 2 \frac{b^3}{a^2} \sqrt{(b^2 - a^2) (a^2+\sqrt{a^2 b^2})}
\nonumber \\
X_2 & = & 0 \nonumber \\
J_1^1 & = & -12 b^4 (1+\frac{b}{a}) \nonumber \\
J_2^1 & = & 0 \nonumber \\
J_1^2 & = & 0 \nonumber \\
J_2^2 & = & -12  b^4 (1+\frac{b}{a}).
\eeqa
The invariance of the vacuum under $SU(2)_D \otimes U(1)_{Ae}$ 
symmetry transformations is
manifested by the fact that $J_1^1 = J_2^2$; 
invariance under $U(1)_Q$
only requires $X_2 = J_1^2 = J_2^1 = 0$, with $J_1^1$ and $J_2^2$
arbitrary.
The scalar potential in this D-flat direction is 
\beqa
\frac{V}{\Lambda_0^4 \lambda^{\frac{18}{11}}} & = &
\frac{1}{2^{12}} \left( \frac{\lambda^{-\frac{1}{11}} \Lambda_0}{b} \right)^{18}
\frac{1}{ \left( 1+ b/a \right)^4} \left( 4 + \frac{8}{1 + b/a}
\right) \nonumber \\ & & 
-\frac{1}{4} \left( \frac{\lambda^{-\frac{1}{11}} \Lambda_0}{b} \right)^{7}
\frac{1}{ \left( 1 + b/a \right)^2} \sqrt{b/a - 1}
\nonumber \\ & & 
+4 \left( \frac{b}{\lambda^{-\frac{1}{11}} \Lambda_0} \right)^{4}
\left( 1 + b/a \right) \left( -1 + 3 b/a \right).
\eeqa
The minimum of the scalar potential is obtained for
\beqa
a & = & 0.5712 \lambda^{-1/11} \Lambda_0 \nonumber \\
b & = & 0.6106 \lambda^{-1/11} \Lambda_0,
\eeqa
and the vacuum energy in the minimum is given by
\beq
V= 2.807 \lambda^{18/11} \Lambda_0^4.
\eeq
Both the expectation values and the vacuum energy agree with
the results as calculated in the full theory \cite{TtV} to lowest
order of  $\lambda/g$,
and in the low energy effective sigma model as described in
the previous section. The method proposed by  Poppitz
and Randall \cite{PR} could, in principle, be used to 
calculate derivatives of the K\"{a}hler potential at this
point in the moduli space, providing yet another method to
calculate the interactions and spectrum in the low energy
theory.

\section{$N_F = 1$}

For $N_F =1$
the global symmetry group of the model is
$SU(3)_F \otimes SU(2)_T \otimes U(1)_{A'} \otimes U(1)_{B'} \otimes U(1)_{R'}$,
under which the fields transform as
\begin{center}
\begin{tabular}{|l|c|c|c|c|c|}
\hline
                    & $SU(3)_{\bar{F}}$ & $SU(2)_T$ &   $U(1)_{A'}$ &   $U(1)_{B'}$   & $U(1)_{R'}$    \\ 
\hline
$\bar{F}^{\alpha}$  &   $3$             &   $1$     &   $ 2$        &   $\frac{1}{3}$ & $-\frac{7}{3}$ \\
$F$                 &   $1$             &   $1$     &   $ 0$        &   $-1$          & $1$            \\
$T_a$               &   $1$             &   $2$     &   $-1$        &   $0$           & $1$            \\
\hline 
\end{tabular} \ \ .
\end{center}
The classical moduli space is described by
the eighteen basic gauge invariants
\beqa
X_{\alpha,a} & = & \epsilon_{\alpha \beta \gamma} \bar{F}_{i}^{\beta}
\bar{F}_{j}^{\gamma} T_{a}^{ij} \nonumber \\
J_{a}^{\alpha} &= & \epsilon_{ijklm} \bar{F}_{n}^{\alpha}
T_a^{ij} T_b^{kl} T_c^{mn} \epsilon^{bc} \nonumber \\
M^{\alpha} & = & \bar{F}_{i}^{\alpha} F^i \nonumber \\
B_{ab} & = & \epsilon_{ijklm} F^i T_a^{jk} T_b^{lm} ,
\eeqa
subject to the two constraints
\beqa
\epsilon_{\alpha \beta \gamma} \epsilon^{ab} J_a^{\alpha} J_b^{\beta} M^{\gamma} 
- \frac{3}{2} \epsilon^{ca} \epsilon^{db} B_{cd} J_a^{\alpha} X_{\alpha,b}
 & = & 0 \label{constr} \\
\epsilon^{ab} J_a^{\alpha} X_{\alpha,b} & = & 0.
\eeqa
The moduli fields transform under the global symmetry group as
\begin{center}
\begin{tabular}{|l|c|c|c|c|c|}
\hline
                    & $SU(3)_{\bar{F}}$ & $SU(2)_T$ &   $U(1)_{A'}$ &   $U(1)_{B'}$   & $U(1)_{R'}$    \\ 
\hline
$X_{\alpha,a}$      &   $\bar{3}$       &   $2$     &   $ 3$        &  $\frac{2}{3}$  &$-\frac{11}{3}$ \\
$J_{a}^{\alpha}$    &   $3$             &   $2$     &   $- 1$       &   $1/3$         &$2/3$           \\
$M^{\alpha}$        &   $3$             &   $1$     &   $2$         &   $-2/3$        &$-4/3$          \\
$B_{ab}$            &   $1$             &   $3$     &   $-2$        &   $-1$          & $3$            \\
\hline 
\end{tabular}\ \ .
\end{center}
There is no
dynamically generated superpotential, but
holomorphy and global symmetries 
allow a modification of the constraint (\ref{constr}) to
\beq
\epsilon_{\alpha \beta \gamma} \epsilon^{ab} J_a^{\alpha} J_b^{\beta} M^{\gamma} 
- \frac{3}{2} \epsilon^{ca} \epsilon^{db} B_{cd} J_a^{\alpha} X_{\alpha,b}
  = \Lambda_1^{b_0},
\eeq
where $b_0 = 10$ for $N_F = 1$, and $\Lambda_1$ is
dynamical scale of the gauge interactions. This modification
is consistent with the requirement that
anomaly matching conditions for unbroken global symmetries
are saturated at points of enhanced symmetry in the moduli space.
Moreover, the modified constraint gives rise to the correct dynamical
superpotential in the $N_F = 0$ model when the additional
flavor is integrated out. To illustrate this holomorphic
decoupling and to study the issue of supersymmetry breaking,
a mass term and a Yukawa term are added to the superpotential:
\beq
W_{tree} =  m_1 M^3 + \lambda X_{3,1}.
\eeq
The constraints can be incorporated into the superpotential by the
introduction of the Lagrange multiplier fields $L_1$ and $L_2$. The
complete superpotential therefore is
\beq
W = W_{tree} + L_1 \left(
\epsilon_{\alpha \beta \gamma} \epsilon^{ab} J_a^{\alpha} J_b^{\beta} M^{\gamma} 
- \frac{3}{2} \epsilon^{ca} \epsilon^{db} B_{cd} J_a^{\alpha} X_{\alpha,b}
-  \Lambda_1^{10} \right) + L_2 \epsilon^{ab} J_a^{\alpha} X_{\alpha,b}.
\eeq
In the  large $m_1$ limit the additional flavor can be integrated out by 
imposing the equations of motion for the fields
$M^{\alpha}$, $J_a^3$, $X_{1,a}$,  $X_{2,a}$ and $B_{ab}$.
With the matching condition
$2 \Lambda_1^{10} m_1 = 9 \Lambda_0^{11}$, the theory for $N_F = 0$ 
with superpotential (\ref{superpotential}) ensues.

Supersymmetry is broken because
the vacuum expectation values of the auxiliary $F$ components
do not all vanish simultaneously.
The equations of motion for the $F$ components are
\beqa
\left( M_{\gamma} \right)_F^{\dagger} =
%\frac{\partial W}{\partial M^{\rho}}
& = & m_1 \delta_{\gamma}^{3}
+ L_1 J_a^{\alpha} J_b^{\beta} \epsilon^{ab} \epsilon_{\alpha \beta \gamma}
\nonumber \\
\left( B^{ef} \right)_F^{\dagger} =
%\frac{\partial W}{\partial B_{ef}} 
& = & -\frac{3}{2} L_1
J_a^{\alpha} X_{\alpha,b} 
\left( \epsilon^{ea} \epsilon^{fb} + \epsilon^{fa} \epsilon^{eb} \right)
\nonumber \\
\left( J_{\alpha}^e \right)_F^{\dagger} =
%\frac{\partial W}{\partial J_e^{\alpha}} 
& = &
2 L_1 J_b^{\beta} M^{\gamma} \epsilon^{eb} \epsilon_{\alpha \beta \gamma}  -
\frac{3}{2} L_1 B_{cd} X_{\alpha,b} \epsilon^{ce} \epsilon^{db} +
L_2 X_{\alpha,b} \epsilon^{eb}
\nonumber \\
\left( X^{\alpha,e} \right)_F^{\dagger} =
%\frac{\partial W}{\partial X_{\alpha,e}} 
& = & 
\lambda \delta_3^{\alpha} \delta_1^e - \frac{3}{2} L_1 B_{cd} J_a^{\alpha}
\epsilon^{ca} \epsilon^{de} + L_2 J_{a}^{\alpha} \epsilon^{ae}
\eeqa
To show that some $F$ fields obtain a vacuum expectation value,
assume first that the vacuum expectation values
of all $F$ components vanish. Then consider
\beq
J_{c}^{\alpha} 
%\frac{\partial W}{\partial M^{\alpha}} 
\left( M_{\alpha} \right)_F^{\dagger} = 
m_1 J_{c}^3.
\eeq
If $<\left( M_{\rho} \right)_F^{\dagger}> = 0$, and $m_1 \neq 0$ then
$<J_c^{3}>=0$.
However, in this case
\beq
%<\frac{\partial W}{\partial X_{3,1}}> 
<\left( X^{3,1} \right)_F^{\dagger}>
= \lambda.
\eeq
This is inconsistent with the assumption. Hence some
$F$ components have a vacuum expectation value, and therefore
supersymmetry is broken.

\section{$N_F = 2$}

For $N_F = 2$ the global symmetry group of the model is
$SU(4)_{\bar{F}} \otimes SU(2)_F \otimes SU(2)_T \otimes U(1)_{A'} 
\otimes U(1)_{B'} \otimes U(1)_{R'}$, and the fields
transform as
\begin{center}
\begin{tabular}{|l|c|c|c|c|c|c|}
\hline
                    & $SU(4)_{\bar{F}}$ & $SU(2)_F$ & $SU(2)_T$ & $U(1)_{A'}$     &   $U(1)_{B'}$   & $U(1)_{R'}$    \\ 
\hline
$\bar{F}^{\alpha}$  &   $4$             &  $1$      &   $1$     & $\frac{3}{2}$   &   $\frac{1}{2}$ & $-\frac{3}{2}$ \\
$F$                 &   $1$             &  $2$      &   $1$     &   $ 0$          &   $-1$          & $1$            \\
$T_a$               &   $1$             &  $1$      &   $2$     &   $-1$          &   $0$           & $1$            \\
\hline 
\end{tabular}\ \ .
\end{center}
The basic gauge invariants which parametrize the classical moduli space
are
\beqa
X_{\alpha \beta,a} & = & \epsilon_{\alpha \beta \gamma \delta} \bar{F}_{i}^{\gamma}
\bar{F}_{j}^{\delta} T_{a}^{ij} \nonumber \\
J_{a}^{\alpha} &= & \epsilon_{ijklm} \bar{F}_{n}^{\alpha}
T_a^{ij} T_b^{kl} T_c^{mn} \epsilon^{bc} \nonumber \\
M_{\sigma}^{\alpha} & = & \bar{F}_{i}^{\alpha} F_{\sigma}^i \nonumber \\
B_{\sigma,ab} & = & \epsilon_{ijklm} F_{\sigma}^i T_a^{jk} T_b^{lm}  \nonumber \\
Y^{\alpha} & = &  \epsilon_{jklmn} \bar{F}_i^{\alpha} T_a^{ij} F_{\sigma}^k F_{\tau}^l T_b^{mn} 
\epsilon^{\sigma \tau} \epsilon^{ab}, \label{moduli}
\eeqa
subject to various constraints. These gauge invariants transform
under the global symmetry transformations as
\begin{center}
\begin{tabular}{|l|c|c|c|c|c|c|}
\hline
                      & $SU(4)_{\bar{F}}$ & $SU(2)_F$ & $SU(2)_T$ & $U(1)_{A'}$      &   $U(1)_{B'}$    & $U(1)_{R'}$    \\ 
\hline
$X_{\alpha \beta,a}$  &   $6$             &  $1$      &   $2$     &   $ 2$           &   $1$            & $-2$           \\
$J_{a}^{\alpha}$      &   $4$             &  $1$      &   $2$     &   $-3/2$         &   $1/2$          & $3/2$          \\
$M_{\sigma}^{\alpha}$ &   $4$             &  $2$      &   $1$     &   $3/2$          &   $-1/2$         & $-1/2$         \\
$B_{\sigma,ab}$       &   $1$             &  $2$      &   $3$     &   $-2$           &   $-1$           & $3$            \\
$Y^{\alpha}$          &   $4$             &  $1$      &   $1$     &   $-1/2$         &   $3/2$          & $5/2$          \\
\hline 
\end{tabular}\ \ .
\end{center}
The model for $N_F=2$ is part of a class of models \cite{CSS} whose
infra--red behavior is commonly referred to as \lq\lq s-confinement \rq\rq. 
The quantum moduli space is identical to the classical one, but its
singularities are interpreted differently. Where in the classical
picture the singularities are associated with  massless
gauge multiplets, at the quantum level they 
are associated with additional massless composite fields.
Close to the origin, all moduli fields (\ref{moduli}) are physical. 
They interact through the confining potential
\beq
W = \frac{1}{\Lambda_2^{b_0}} \left( -JJMM + 3 JYX + 3 JBXM - \frac{9}{32} BBXX \right),
\eeq
where $b_0 = 9$ for $N_F = 2$, and
\beqa
JJMM & = & J_a^{\alpha} J_b^{\beta} M_{\sigma}^{\gamma} M_{\tau}^{\delta} 
           \epsilon_{\alpha \beta \gamma \delta} \epsilon^{\sigma \tau} \epsilon^{ab} \nonumber \\
JYX & = & J_a^{\alpha} Y^{\beta} X_{\alpha \beta, b} \epsilon^{ab} \nonumber \\
JBXM & = & J_a^{\alpha} B_{\sigma,cd} X_{\alpha \beta,b} M_{\tau}^{\beta} 
           \epsilon^{ac} \epsilon^{bd} \epsilon^{\sigma \tau} \nonumber \\
BBXX & = & B_{\sigma,ab} B_{\tau,cd} \epsilon^{\sigma \tau} \epsilon^{ac}
           X_{\alpha \beta,e} X_{\gamma \delta, f} 
           \epsilon^{\alpha \beta \gamma \delta} \epsilon^{be} \epsilon^{df}.
\eeqa
The coefficients of the terms in this potential are such that
the F-flatness conditions reproduce 
the constraints of the classical moduli space.
At the origin none of the global symmetries are broken. All anomaly
coefficients in the fundamental theory and the low energy
effective theory match, providing a stringent test for this picture.
In order to show that the model for $N_F=2$ is related to
the model for $N_F = 1$ by holomorphic decoupling and to
study dynamical supersymmetry breaking, the superpotential
\beq
W_{tree} = m_1 M_1^3 + m_2 M_2^4 + \lambda X_{34,1}
\eeq
is added.
In the limit $m_2 >> \Lambda_2$ the heavy degrees of freedom
can be integrated out, and the theory for $N_F = 1$ results
with the matching condition $m_2 \Lambda_2^9 = 2 \Lambda_1^{10}$. Supersymmetry
is broken by the O'Raifeartaigh mechanism. The proof
is by reductio ad absurdum. Assume that none of the
auxiliary $F$ fields has an expectation value. Note that
\beq
J_a^{\alpha} \left( M_{\alpha}^{\sigma} \right)_F^{\dagger} - 
\left( Y_{\alpha} \right)_F^{\dagger} J_c^{\alpha} B_{\tau,da} \epsilon^{cd} \epsilon^{\tau \sigma} 
 =  m_1 J_a^3 \delta_1^{\sigma} +  m_2 J_a^4 \delta_2^{\sigma}.
\eeq
Therefore, if $<(M_a^{\alpha})_F^{\dagger}> = 0$ and
$<(Y_{\alpha})_F^{\dagger}> = 0$, then $<J_a^3> =0$ and $<J_a^4 >= 0$
for $m_1 \neq 0$ and $m_2\neq 0$.
In addition,
\beq
( M_{\alpha}^1 )_F^{\dagger} X_{\beta \gamma,a} \epsilon^{\alpha \beta \gamma 4} -
( Y_{\alpha})_F^{\dagger} X_{\beta \gamma,c} B_{\tau,ab} \epsilon^{bc} 
\epsilon^{\alpha \beta \gamma 4} \epsilon^{\tau 1} +
\frac{8}{3} (Y_{\alpha})_F^{\dagger} J_a^{\alpha} M_{\tau}^4 \epsilon^{\tau 1}
= -2 m_1 X_{12,a}.
\eeq
As a consequence, $<X_{12,a}> = 0$ if $<( M_{\alpha}^1 )_F^{\dagger}>=0$
and $<( Y_{\alpha})_F^{\dagger}>=0$. However, in that
case $<(X^{\alpha \beta,a})_F^{\dagger}>=\lambda$, in contradiction with 
our assumption. Some $F$ components therefore have an expectation value,
and supersymmetry is broken.

\section*{Conclusions}

The classical moduli space for the model with $N_F = 0$ without
a superpotential has a bigger 
symmetry group than the fundamental theory.
In the present work this followed algebraically from
the explicit calculation of the K\"{a}hler potential in the
moduli space. However, it seems to us that  the extended
symmetry of the moduli space is probably a necessary consequence of
supersymmetry and the specific representation of the matter fields
under the gauge group. We surmise therefore that there must be
a more elegant method based on representation theory and
geometry to determine the symmetries of the classical moduli space in this
particular model, and perhaps also in supersymmetric gauge theories
in general. We may attempt to address this issue in a later
paper.

The superpotential in the model with $N_F=0$ is invariant
under part of the extended symmetries of the effective K\"{a}hler
potential in the moduli space. Even after spontaneous symmetry
breaking in the effective low energy sigma model some of the
extended symmetry remains. This remnant explains 
previously observed degeneracies
in the mass spectrum. It is interesting to note that even
though the symmetry breaking patterns as displayed in the
fundamental theory and the effective low energy sigma model
are different, the number of broken generators is identical
in both cases. The number of Goldstone bosons is therefore
also the same, which is, of course, required for consistency.
All the features of this calculable model are
now well understood. It can therefore be used as a controlled
laboratory for dynamical supersymmetry breaking, just as the 
$SU(3) \otimes SU(2)$ model.

For the calculable limit of the $N_F = 0$ model
the existence of a supersymmetric vacuum at the origin at strong coupling
cannot strictly be excluded.
In Ref. \cite{ADS2} a strong case was made
that such behavior is implausible. Moreover, even if such
a supersymmetric vacuum existed, the supersymmetry breaking
vacuum at weak coupling would be at least meta--stable.
As the $N_F = 1$ and $N_F=2$ models are strongly interacting,
the argument that supersymmetry is broken in these models
hinges upon the assumed correct identification of the
low energy degrees of freedom and the K\"{a}hler potential.
Although the arguments for broken supersymmetry are
quite convincing in each case, there always remains a loophole. 

Further support for the hypothesis that supersymmetry is
broken is provided by the following argument\cite{IT,Mur}:
The models we consider are related by holomorphic decoupling. i.e. a model
with less flavors can be obtained by varying mass parameters
of a model with more flavors.
It was shown in Ref. \cite{HS} that if supersymmetry is broken
for a range of the parameters, it is broken for generic values of
those parameters, with the possible exception of isolated special 
points.
It then follows that if supersymmetry is broken in one
of the models under consideration, it is broken in all of them.

\section*{Acknowledgements}  
The author thanks Tom Clark, Tom Kephart and Xerxes Tata
for useful discussions. This project was started
at Vanderbilt University. The hospitality of
Fermilab and Brookhaven National Laboratory, where part of the work
was done, is gratefully acknowledged. 
This work was supported in part by the U.S. Department
of Energy under grants No. DE-FG05-85ER40226 and No. DE-FG03-94ER40833.

\end{document}